\documentclass[12pt]{article}
\usepackage{epsfig, amsmath, amssymb}
\usepackage{graphicx,psfrag} 
\newcommand{\be}{\begin{equation}}
\newcommand{\ee}{\end{equation}}
\newcommand{\bea}{\begin{eqnarray}}
\newcommand{\eea}{\end{eqnarray}}
\newcommand{\bA}{\begin{array}}
\newcommand{\eA}{\end{array}}
\newcommand{\bc}{\begin{center}}
\newcommand{\ec}{\end{center}}
\newcommand{\al}{\alpha}

\newcommand{\ra}{\rightarrow}
\newcommand{\del}{\partial}

\newcommand{\ie}{{\it i.e.}}
\newcommand{\eg}{{\it e.g.}}

\newcommand{\Nf}{${\cal N}{=}4$}

\begin{document}

%\ifeprint
%\fi

\begin{titlepage}
\vspace{20mm}

\bc

\hfill  {TIFR/TH/09-12} \\
\hfill  {\tt arXiv:0904.4532 [hep-th]} \\
         [22mm]
%%X\vfill

{\huge String spectra near\\ [2mm]
some null cosmological singularities}
\vspace{16mm}

{\large Kallingalthodi Madhu$^a$ and K.~Narayan$^b$} \\
\vspace{3mm}
{$^a$\small \it Tata Institute of Fundamental Research,\\}
{\small \it Homi Bhabha Road, Colaba, Mumbai 400005, India.\\}
\vspace{2mm}
{$^b$\small \it Chennai Mathematical Institute, \\}
{\small \it SIPCOT IT Park, Padur PO, Siruseri 603103, India.\\}
%{\small Email: \ narayan@cmi.ac.in}\\

\ec
\medskip
\vspace{25mm}

\begin{abstract}
We construct cosmological spacetimes with null Kasner-like
singularities as purely gravitational solutions with no other
background fields turned on. These can be recast as anisotropic
plane-wave spacetimes by coordinate transformations. We analyse string
quantization to find the spectrum of string modes in these
backgrounds.  The classical string modes can be solved for exactly in
these time-dependent backgrounds, which enables a detailed study of
the near singularity string spectrum, (time-dependent) oscillator 
masses and wavefunctions. We find that for low lying string
modes (finite oscillation number), the classical near-singularity 
string mode functions are non-divergent for various families 
of singularities. Furthermore, for any infinitesimal regularization 
of the vicinity of the singularity, we find a tower of string modes 
of ultra-high oscillation number which propagate essentially freely 
in the background. The resulting picture suggests that string 
interactions are non-negligible near the singularity.
\end{abstract}

\end{titlepage}

\newpage 
{\footnotesize
\begin{tableofcontents}
\end{tableofcontents}
}

\vspace{2mm}

\section{Introduction}

Understanding cosmological singularities in string theory is an
important goal, and has been the subject of several investigations
\eg\ \cite{horowitzsteif,bhkn,cornalbacosta,lms,lawrence,horopol,
ckr,prt,david,bbop,grs,bp0307,bdpr,matrixmodelsCosmo,
McGreevy:2005ci,csv0506,li0506,blauMpp,cosmoMatrix,Silverstein:2005qf,
keshav0302,hertoghoro,dmnt,adnnt,Chu:2006pa,linwen,
Turok:2007ry,crapsetal,crapsetal2}.

Our work in this paper has been in part motivated by investigations 
\cite{dmnt,adnnt} involving generalizations of AdS/CFT where the bulk 
contains null or spacelike cosmological singularities, with a nontrivial 
dilaton $e^\Phi$ that vanishes at the location of the cosmological 
singularity, the curvatures behaving as\ $R_{MN}\sim\del_M\Phi\del_N\Phi$. 
The gauge theory duals are \Nf\ \ Super Yang-Mills theories with a 
time-dependent gauge coupling $g_{YM}^2=e^\Phi$, and \cite{dmnt,adnnt} 
describe aspects of the dual descriptions of the bulk cosmological
singularities. From the bulk point of view, supergravity breaks down
and possible resolutions of the cosmological singularity stem from
stringy effects. Indeed noting $\al'\sim{1\over g_{YM}^2N}$ from the
usual AdS/CFT dictionary and extrapolating naively to these time-dependent
cases with a nontrivial dilaton, we have\ $\al'\sim{1\over e^\Phi N}$\
indicating vanishing effective tension for stringy excitations, when\
$e^\Phi\ra 0$\ near the singularity. While this is perhaps wrong in 
detail, we expect that stringy effects are becoming important near 
the bulk singularity, corresponding to possible gauge coupling effects 
in the dual gauge theory. It is therefore interesting to understand
worldsheet string effects in the vicinity of the singularity. Owing to
the technical difficulties with string quantization in an AdS
background with RR flux, we would like to look for simpler, purely
gravitational backgrounds as toy models whose singularity structure
shares some essential features with the backgrounds in the AdS/CFT
investigations.  We first find such spacetime backgrounds as
``near-singularity'' solutions to Type II supergravity (preserving a
fraction of lightcone supersymmetry). In general, these are null
Kasner-like solutions with null cosmological singularities (we also
find approximate solutions that extrapolate from these
near-singularity solutions to flat space asymptotically). These can be
recast as anisotropic plane-wave-like spacetimes by a coordinate
transformation, and we outline arguments in these coordinates
suggesting the absence of higher derivative curvature corrections to
these spacetimes (which are essentially plane-wave backgrounds).

We then perform an analysis of string quantization to find the
spectrum of string modes in these backgrounds. We find it convenient
to use (Rosen-like) coordinates where the null cosmology
interpretation is manifest. With these lightlike backgrounds, it is
natural to use lightcone gauge. The classical string modes can be
solved for exactly in these time-dependent backgrounds, which enables
a detailed study of the near singularity string spectrum. For various
families of singularities, the classical string oscillation amplitudes
for low lying oscillation number $n$ are non-divergent near the
singularity, with asymptotic time-dependence similar to the
center-of-mass modes.  From the Hamiltonian, we find time-dependent
masses for these string oscillator modes. However, for any
infinitesimal regularization of the vicinity of the singularity, say
$\tau\lesssim\tau_\epsilon$, we find string modes of ultra-high
oscillation number $n\gg {1\over\tau_\epsilon^{a+1}}$ which propagate
essentially freely in the background. The near-singularity region thus
appears to be filled with such highly stringy modes. There have been
several investigations of string quantization in plane-wave
backgrounds with singularities \cite{prt,david,bbop,crapsetal,
  crapsetal2} and our string analysis has some overlap with \cite{prt}
in particular.

In sec.~2, we describe the spacetime backgrounds. Sec.~3 describes the
string quantization. Sec.~4 contains a discussion and open questions.
In Appendix A, we describe some properties of the spacetime backgrounds, 
while Appendix B outlines string quantization in coordinates
corresponding to a different time parameter.

\section{The spacetime backgrounds}

We are interested in spacetime backgrounds that have a Big-Bang or 
Big-Crunch type of singularity at some value of the lightlike time 
coordinate $x^+$. We also want to restrict attention to purely 
gravitational solutions for simplicity, \ie\ with unexcited dilaton and 
RR/NSNS fields. This means we want to solve the equations $R_{MN}=0$. 
Null-time dependence reduces these equations to $R_{++}=0$. 

Let us begin by considering a spacetime background with two scale factors, 
of the form
\be
ds^2 = g_{\mu\nu} dx^\mu dx^\nu = 
e^{f(x^+)}\left(-2dx^+dx^- + dx^idx^i\right) + e^{h(x^+)}dx^mdx^m\ ,
\ee
where $i=1,2$, $m=3,\ldots,D-2$. One may also think of the $x^m$ directions 
as compactified, representing say a $T^{D-4}$. For the critical superstring 
with $D=10$, we could alternatively replace this 6-dim transverse space by 
some Ricci flat space such as a Calabi-Yau 3-fold. The intuition here is 
that the time-dependence of the ``internal'' space induces time-dependence 
in the 4-dim spacetime as well, as in \eg\ \cite{townwohl,TimeMsolns}.
Another perspective is that the internal space scale factor is the 
analog of the dilaton in the AdS/CFT cosmological solutions of 
\cite{dmnt,adnnt}, as we will elaborate on below.

Simple classes of singularities in this system are obtained for spacetimes 
whose limiting form in the vicinity of $x^+=0$ is null 
Kasner-like\footnote{A spacetime of the form (\ref{absolns}) 
(\ref{absolns2}) with $a<0$ can be transformed by a change of coordinates 
to one with $a>0$ by redefining\ $y^+={1\over x^+}$. This recasts\ 
$g_{+-}=(x^+)^{-|a|}=(y^+)^{|a|}$\ and moves the singularity at 
$x^+\ra\infty$ in the spacetimes with $a<0$ to $y^+=0$. Thus it is 
sufficient to study spacetimes (\ref{absolns}) (\ref{absolns2}) with $a>0$.}
\be\label{absolns}
ds^2 = (x^+)^a\left(-2dx^+dx^-+dx^idx^i\right) + (x^+)^b dx^mdx^m\ ,
\qquad\ \ a>0\ .
\ee
More generally, consider spacetimes of a general Kasner-like form
\be\label{absolns2}
ds^2 = (x^+)^a\left(-2dx^+dx^- + dx^idx^i\right) + (x^+)^{b_m} dx^mdx^m\ ,
\qquad\ \ a>0\ ,
\ee
\ie\ the individual internal dimensions $x^m$ evolve independently according 
to their Kasner exponents $b_m$ appearing in the individual scale factors 
$e^{h_m(x^+)} \ra\ (x^+)^{b_m}$ as $x^+\ra 0$.

The coordinate transformation\ $x^I=(x^+)^{-a_I/2} y^I$, where 
$a_I\equiv a,b_m$,\ gives
\be
(x^+)^{a_I}(dx^I)^2 = (dy^I)^2 - {a_Idx^+y^Idy^I\over x^+} 
+ {a_I^2(y^I)^2 (dx^+)^2\over 4(x^+)^2}\ .
\ee
Then the metric (\ref{absolns2}) becomes of manifest plane-wave form
\be\label{planewave}
ds^2 = -2(x^+)^a dx^+dy^- + 
\left[\sum_I \left({a_I^2\over 4}-{a_I(a+1)\over 2}\right) (y^I)^2\right] 
{(dx^+)^2\over (x^+)^2} + (dy^I)^2\ ,
\ee
where we have redefined\ $y^-=x^-+({\sum_Ia_I(y^I)^2\over 4(x^+)^{a+1}})$. 
For $a_I=a,b_m$ distinct, these are in general anisotropic plane-waves 
with singularities (after further redefining $(x^+)^adx^+=d\lambda$). 
In what follows, we will find it convenient to work in the (Rosen) 
coordinates (\ref{absolns}), (\ref{absolns2}), where the null cosmology 
interpretation is manifest, but as we will see below, there are 
close parallels with various previous studies on plane-wave spacetimes 
with singularities, most notably \cite{prt} (see also 
\cite{david,crapsetal,crapsetal2}).

The spacetimes (\ref{absolns}) have nonvanishing Riemann curvature 
components (with \eg\ $f'\equiv {df\over dx^+}$)
\bea
&& R_{+i+i} = {1\over 4}\left((f')^2-2f''\right)e^{f(x^+)} 
= {a(a+2)\over 4} (x^+)^{a-2}\ ,\nonumber\\
&& R_{+m+m} = {1\over 4}\left(2f'h_m'-(h_m')^2-2h_m''\right)e^{h_m(x^+)} 
= {b(2a+2-b)\over 4} (x^+)^{b-2}\ ,
\eea

For these spacetimes to be Ricci-flat solutions of the Einstein equations, 
the equation of motion\ $R_{++}=0$\ must hold, giving
\bea\label{EOMfhm}
&& R_{++}={1\over 2}(f')^2-f''+{1\over 2}\sum_m (-2h_m''-(h_m')^2+2f'h_m') = 0
\ \nonumber\\
&& \ \ \Rightarrow\ \qquad
a^2+2a + {1\over 2} \sum_m (-b_m^2 + 2b_m + 2ab_m) = 0\ .
\eea
This relates the various (null) Kasner-like exponents $a, b_m$. 
The equation in terms of the general scale factors shows that the 
curvature for the 4D scale factor $e^f$ is sourced by those for the 
``internal'' scale factors $e^{h_m}$: indeed the $h_m$ are the analogs 
of the dilaton scalar in the AdS/CFT cosmological context 
\cite{dmnt,adnnt} where the corresponding equation was\ 
$R^{(4)}_{++}={1\over 2} (\del_+\Phi)^2$. That is, the kinetic terms\ 
$(\del_+h_m)^2$\ (and related cross-terms) play the role of the 
dilaton in driving the singular behaviour of the 4D part of the spacetime.

In what follows, we will specialize to the symmetric case here, \ie\ all 
$b_m\equiv b$ equal (and $e^{h_m}\equiv e^h$). 
Then $R_{++}=0$ simplifies to
\be\label{EOMfh}
{1\over 2}(f')^2-f''+{D-4\over 2} (-2h''-(h')^2+2f'h') = 0
\ \Rightarrow\ \ a^2+2a+{D-4\over 2}(-b^2+2b+2ab) = 0\ .
\ee
If $b=a$, this equation (assuming $D>2$) simplifies to give the solutions\ 
$b=a=0,-2$, in which case the Riemann curvature components are seen to 
identically vanish (the solution $(-2,-2)$ can be shown to be flat space 
by the coordinate transformation to plane-wave form). Thus an interesting 
solution requires that the ``internal'' $x^m$-space either grows or 
shrinks faster than the spatial part of the 4-dim cosmology. For any 
$b\neq a$, the equation of motion above is a quadratic in $a$ that admits 
various solutions with
\be\label{abvalues}
2a=-2-(D-4)b\pm\sqrt{4+(D-4)(D-2)b^2}\ .
\ee
Taking the positive radical, it can be seen that restricting $a>0$ for our 
solutions implies\ $b>2$ or $b<0$. Furthermore\ $a+1-b>0$ if\ $b<0$\ or\ 
$|b|<{\sqrt{2}\over D-2}$.

Suppose we focus on finding solutions with $a,b$, being even integers, 
so that the metric allows unambiguous analytic continuation from 
$x^+<0$ to $x^+>0$ across the singularity. One may imagine that this 
is a coordinate-dependent choice of the time parameter $x^+$ and therefore 
not sacrosanct: however if we do take this choice, $a,b$ being even 
integers seems natural.
This is more restrictive: we must consider the above as Diophantine 
quadratic equations with solutions over integers, which are in general 
rarer. We then need to look for those $b$ for which the radical above is 
integral. For the cases of obvious interest, \ie\ the bosonic string 
($D=26$) and the superstring ($D=10$), the radicals simplify to 
$2\sqrt{1+132b^2}$ and $2\sqrt{1+12b^2}$ respectively. It is then 
straightforward to check that 
\bea
&& (a,b) = (0,2), (44,-2), (44,92), (2068,-92) \ldots \qquad\qquad\qquad 
[D=26]\ , \nonumber\\
&& (a,b) = (0,2),(12,-2), (12,28), (180,-28), (180,390),
\ldots \qquad  [D=10]\ , 
\eea
are solutions. Our analysis of these solutions in what follows will not 
depend on these detailed values though.

These solutions can be thought of as the ``near-singularity'' limiting
regions of more general spacetimes where the scale factors $e^f,
e^{h_m}$ are not necessarily of power-law type\footnote{These solutions 
also arise as certain Penrose limits starting with some cosmological 
spacetimes and adding a spectator dimension \cite{prt} (see also 
\cite{blauPenrose}).}. Since the various scale factors $e^f, e^{h_m}$, 
are related by the single equation of motion (\ref{EOMfhm}), a generic
choice of $e^f$ admits a solution to (\ref{EOMfhm}) for the remaining
scale factors $e^{h_m}$.  For instance, with a single scale factor
$e^{h_m}=e^h$,\ taking\ $e^f=\tanh^a(x^+)$, we can in principle solve
for $e^h$. In the limiting near-singularity region, we have already
seen null-Kasner-like solutions with (\ref{EOMfh}) relating the Kasner
exponents. In the asymptotic region of large $x^+$, it can be checked
that
\be\label{asympsoln}
e^f=\tanh^a(x^+)\ra\ 1-2ae^{-2x^+}\ , \qquad
e^h\sim {\rm const} + {2a\over D-4} e^{-2x^+}\ ,
\ee
is an approximate solution to (\ref{EOMfh}) (dropping the subleading 
nonlinear terms).

We now make a few comments on the cosmological singularities in these
spacetimes. No curvature invariants diverge due to the lightlike nature 
of this system, since no nontrivial contraction is nonzero. However
there are diverging tidal forces for null geodesic congruences. 
Consider for instance a simple class of null geodesic congruences 
propagating solely along $x^+$ (at constant $x^-,x^i,x^m$), with 
cross-section along the $x^i$ or $x^m$ directions. These are described 
by ($\Gamma^+_{++}=f'={a\over u}$ is the only nonzero $\Gamma^+_{ij}$)
\be
{d^2x^+\over d\lambda^2} + \Gamma^+_{ij} \left({dx^i\over d\lambda}\right) 
\left({dx^j\over d\lambda}\right) = {d^2x^+\over d\lambda^2} 
+ \Gamma^+_{++} \left({dx^+\over d\lambda}\right)^2 = 0\ .
\ee
This gives the affine parameter along these null geodesics 
\be\label{affineparam}
\lambda = const. \int dx^+ e^{f(x^+)} = const. \int dx^+ (x^+)^a 
= const. {(x^+)^{a+1}\over a+1}\ .
\ee
and the tangent vector 
\be
\xi = \del_{\lambda} = \left({dx^+\over d\lambda}\right)\del_+ 
\equiv \xi^+\del_+\ .
\ee
The relative acceleration of neighbouring geodesics in a null congruence 
can be calculated using the geodesic deviation equation giving
\be
a^M = g^{MN} R_{NCBD} \xi^C \xi^D n^B
\ee
where $n=n^B\del_B$ is the separation vector along a cross-section of 
the congruence. For our system, this gives
\be
a^i = g^{ii} R_{+i+i} (\xi^+)^2 n^i = {a(a+2) n^i\over 4 (x^+)^{2a+2}}\ ,\qquad
a^m = g^{mm} R_{+m+m} (\xi^+)^2 n^m = {b(2a+2-b) n^m\over 4 (x^+)^{2a+2}}\ .
\ee
The corresponding invariant acceleration norms are
\be
|a^i|^2 = g_{ii}a^ia^i\sim  {1\over (x^+)^{3a+4}}\ , \qquad
|a^m|^2 = g_{mm}a^ma^m\sim {1\over (x^+)^{-b+4a+4}}\ ,
\ee
So we see diverging tidal forces as $x^+\ra 0$ for spacetimes satisfying 
the conditions (restricting to $a>0$)
\be\label{abCondns}
b<4a+4\ , \qquad  a>0\ ,
\ee
indicating a singularity\footnote{This is true except when the 
coefficients of all $a^I$ vanish:\ this happens for the spacetimes 
$(a,b)=(0,0), (0,2)$.}. Since tidal forces diverge (for $a,b$, satisfying 
both these conditions) along both the $x^i$ and the $x^m$ directions, 
the locus of the singularity is the 8-dim space spanned by the $x^i,x^m$. 
From the point of view of a Penrose-like diagram, we see that the 
singularity locus extends all the way to $x^-\ra\infty$. 
We will see reflections of this later in the string worldsheet analysis.

In Appendix A, we show that these spacetime backgrounds preserve 16 
real (lightcone) supercharges. This is not a feature we use however, 
and our worldsheet analysis below does not appear to depend crucially 
on spacetime supersymmetry of these backgrounds.

We also mention that these spacetimes appear to not admit $\al'$ 
corrections due to higher order curvature terms, as is often the case 
with lightlike backgrounds. This is perhaps not surprising in light of 
the coordinate transformation that casts these null cosmologies in the 
form of anisotropic plane waves, which are known to be devoid of higher 
derivative corrections.

In general, these spacetimes are slightly different from those studied
by \eg \cite{bhkn,cornalbacosta,lms,lawrence,horopol} which were
time-orbifold-like spacetimes. Although there are conceptual 
similarities, the detailed structure of the spacetimes are different
and in particular, there is no issue of backreaction due to several
orbifold ``images'' \cite{lawrence, horopol}.

In what follows, we analyse the string spectrum in the vicinity of 
the cosmological singularities of these spacetimes.

\section{A string worldsheet analysis}

We will now describe a worldsheet analysis of string propagation in 
these backgrounds. Consider the worldsheet action for the closed string 
propagating in such backgrounds
\be
 S = -\frac{1}{4\pi\al'} \int d\tau d\sigma\ \sqrt{-h} h^{ab}\ 
\del_a X^\mu \del_b X^\nu g_{\mu\nu}(X)
\ee
The worldsheet metric $h_{ab}$ has the signature (-1,1). It is convenient 
in the worldsheet analysis to use lightcone gauge\ $x^+=\tau$, in keeping 
with the null structure of the spacetimes in question here. Unlike flat 
space however, it is not possible in general to use both lightcone gauge 
$x^+=\tau$ and conformal gauge\ $h_{ab}\propto\eta_{ab}$ since that is one 
gauge condition too many, as we will see below. Let us therefore begin 
by setting\ $h_{\tau\sigma}=0$,\ to simplify the worldsheet action, as in 
\cite{Polchinski:2001ju} (see also 
\cite{Metsaev:2000yf})\footnote{Ref.~\cite{blauBrinkman} studies some 
aspects of string quantization in Brinkman coordinates.}. 
Then the worldsheet Lagrangian becomes
\be
{\cal L} = -{1\over 4\pi\al'} \int d\sigma\ 
\left( -E g_{IJ} \del_\tau X^I \del_\tau X^J 
+ {1\over E}\ g_{IJ} \del_\sigma X^I \del_\sigma X^J 
- 2E g_{+-} \del_\tau X^- \right)\ ,
\ee
where we have defined\ 
$E(\tau,\sigma)=\sqrt{-{h_{\sigma\sigma}\over h_{\tau\tau}}}$.
Since $X^-$ is not dynamical, we can eliminate this and reduce the system 
to the physical transverse degrees of freedom. Now if $E=1$ is allowed, 
then we have\ $h_{\tau\tau}=-h_{\sigma\sigma}$, which is equivalent to 
conformal gauge being compatible with lightcone gauge. However, since the 
momentum conjugate to $X^-$ is\ \ $p_-={E g_{+-}\over 2\pi\al'}$\ which is 
a $\tau$-independent constant, we have\ $E=-{1\over g_{+-}}$ (setting 
$p_-=-{1\over 2\pi\al'}$ by a $\tau$-independent reparametrization 
invariance). Thus we see that conformal gauge is 
disallowed\footnote{Appendix B contains a discussion with the affine 
parameter $\lambda$ being the time parameter: in this case, $g_{+-}=-1$, 
and conformal gauge is compatible with lightcone gauge.} since 
$g_{+-}\neq -1$. The action for our background simplifies to
\be\label{action}
S = {1\over 4\pi\al'} \int d^2\sigma\ \left( (\del_\tau X^i)^2 - e^{2f(\tau)} 
(\del_\sigma X^i)^2 + e^{h(\tau)-f(\tau)} (\del_\tau X^m)^2
- e^{h(\tau)+f(\tau)} (\del_\sigma X^m)^2 \right)
\ee
This action contains only the physical transverse oscillation modes\ 
$X^I\equiv X^i, X^m$, of the string. In effect, all the gauge freedom 
and corresponding constraints have been used up, with\ $X^-=x^-_0+p_-\tau$.

The corresponding Hamiltonian, $-p_+$, re-expressing the momenta $\Pi^I$ 
in terms of $\del_\tau X^I$, is
\be\label{Hamiltonian}
H = {1\over 4\pi\al'} \int d\sigma \left[ (\del_\tau X^i)^2 
+ e^{2f(\tau)} (\del_\sigma X^i)^2 + e^{h(\tau)-f(\tau)} (\del_\tau X^m)^2 
+ e^{h(\tau)+f(\tau)} (\del_\sigma X^m)^2 \right]
\ee
In general, one might imagine that a time-dependent background pumps in
energy and excites string modes, and the classical Hamiltonian above does 
reflect this. For spacetimes satisfying\
$e^{2f}\ra 0$\ near the singularity\ $\tau\ra 0$, the potential energy 
of the $X^I$ modes due to the $e^{2f(\tau)}$ factor becomes vanishingly 
small near $x^+=0$ (for $a>0$). This could be taken to mean that it 
costs vanishingly little energy to create long strings as we approach 
$x^+=\tau=0$, the effective tension of string modes becoming vanishingly 
small near the singularity. However this appears to be misleading: 
what is relevant is \eg\ the ratio\ 
${e^{2f} (\del_\sigma X^i)^2\over (\del_\tau X^i)^2}$. This has a more 
detailed form involving nontrivial $\tau$-dependence stemming from 
both $g_{IJ}$ and from the asymptotic behaviour of string modes $X^I$, 
which we can solve for exactly in this background. Furthermore since this
Hamiltonian corresponds to $x^+$-translations and $g_{+-}\neq -1$, the 
string oscillator masses, which are coordinate invariant, are\ 
$m^2\sim g^{+-}p_+p_-$, whose $\tau$-dependence is different from that 
of the Hamiltonian. In the case of affine parameter quantization 
(Appendix B), the time-dependence of the Hamiltonian translates directly 
to that of the oscillator masses.

Heuristically one might imagine that the string gets highly excited
and breaks up into bits propagating independently near the singularity: 
in a sense, this is akin to a worldsheet analog of the observations of
BKL \cite{BKL} on ultralocality near a cosmological singularity. It
would be interesting to understand this better. We will find some
parallels with this in our analysis later, which will reveal
distinctly stringy behaviour.

In the next section, we will study quantum string propagation in this
background in detail. We will focus on the symmetric case, \ie\ all
$b_m=b$ equal giving two exponents $a,b$, but it is straightforward to
generalize our analysis to the general case.

\subsection{String modes and quantization}

We are interested in studying the behaviour of string modes as we 
approach the singularity from the past, \ie\ $\tau<0$. For notational 
convenience, we will simply use $\tau$ to denote $|\tau|=-\tau$ in the 
expressions below.
The equations of motion from the worldsheet action above are
\bea
\label{EOM1}
\del_\tau^2 X^i - e^{2f(\tau)} \del_\sigma^2 X^i = 0\ ,\qquad\ \nonumber\\
\del_\tau^2 X^m  + (\del_\tau h - \del_\tau f) 
\del_\tau X^m - e^{2f(\tau)} \del_\sigma^2 X^m = 0\ ,
\eea
which simplify in the near-singularity region of spacetime to
\bea
\label{EOM2}
\del_\tau^2 X^i - \tau^{2a} \del_\sigma^2 X^i = 0\ , \nonumber\\
\del_\tau^2 X^m  + {b-a\over \tau}\ \del_\tau X^m -
\tau^{2a} \del_\sigma^2 X^m = 0
\eea
Decomposing the $X^I$ as\ $f^I_n(\tau) e^{in\sigma}$,\ we can show that the 
time-dependent mode solutions of these equations are given in terms of 
arbitrary linear combinations of two Bessel functions\footnote{Setting\ 
$f^i_n\ra\sqrt{\tau}f^i_n ,\ f^m_n\ra\tau^\nu f^m_n$,\ transforms the 
equations of motion (\ref{EOM2})\ to the standard Bessel forms
\bea
t^2 {f^i_n}'' + t {f^i_n}' + (t^2 - {1\over 4 (a+1)^2}) f^i_n = 0\ , \qquad\ \
t^2 {f^m_n}'' + t {f^m_n}' + (t^2 - {\nu^2\over (a+1)^2}) f^m_n = 0\ ,
\qquad\ t={n\tau^{a+1}\over a+1}\ .
\nonumber
\eea
}
\bea\label{modesolns}
f^i_n(\tau) =  {c^i_{n1}} \sqrt{n \tau}\ 
J_{\frac{1}{2a+2}}\left(\frac{n \tau^{a+1}}{a+1} \right) + {c^i_{n2}} 
\sqrt{n \tau}\ Y_{\frac{1}{2a+2}}\left(\frac{n \tau^{a+1}}{a+1} \right)\ ,
\qquad\qquad\ \nonumber\\
f^m_n(\tau) = {c^m_{n1}}\sqrt{n}\ 
\tau^\nu\ J_{{\nu\over a+1}}\left({n \tau^{a+1}\over a+1}\right)
+ {c^m_{n2}}\sqrt{n}\ \tau^\nu\ 
Y_{{\nu\over a+1}}\left({n \tau^{a+1}\over a+1}\right) ,
\qquad \nu={a+1-b\over 2}\ .
\eea
These expressions are valid for $\nu>0$, while similar Bessel functional 
forms with index ${|\nu|\over a+1}$ hold for $\nu<0$. The Bessel index in
$f^i_n$ is thus always less than ${1\over 2}$ since $a>0$, while for 
$b<0$, the Bessel index in $f^m_n$ is always greater than ${1\over 2}$.
The complex coefficients $c^I_{n1},c^I_{n2}$ can be taken to indicate 
the choice of a vacuum by defining positive/negative frequency modes.
For now, we keep them as two independent unfixed constants: we will 
comment on specific choices at appropriate points in what follows.

Note the similarity between these string worldsheet mode solutions and 
the well-known Hankel function description of spacetime scalar modes 
propagating in 4-dim de Sitter backgrounds. Spacetime scalar modes in 
the present null Kasner-like backgrounds are somewhat different however 
from these\footnote{Consider a massive scalar $\phi$ in the background 
(\ref{absolns}), with action\ 
$S=\int d^Dx \sqrt{-g} (-g^{\mu\nu} \del_\mu\phi\del_\nu\phi - m^2\phi^2)$ ,
and equation of motion\ 
${1\over\sqrt{-g}} \del_\mu(\sqrt{-g} g^{\mu\nu} \del_\nu\phi) - m^2\phi = 0$.\
Taking modes\ $\phi=f(x^+) e^{ik_-x^-+ik_ix^i+ik_mx^m}$,\ this simplifies to\
${1\over f} {df\over dx^+} 
= {i\over 2k_-}\left( k_i^2 + k_m^2 (x^+)^{a-b} + m^2 (x^+)^a 
+ {2a+(D-4)b\over 2x^+} \right) ,$
\ which can be solved to give
\bea
\phi(x^\mu) = {\rm exp}\left[{i\over 2k_-} \left( k_i^2 x^+ 
+ k_m^2 {(x^+)^{a+1-b}\over a+1-b} + m^2 {(x^+)^{a+1}\over a+1} 
+ {2a+(D-4)b\over 2} \log x^+ \right)\right]\ .\nonumber
\eea
Thus generically these modes have a phase that oscillates ``wildly'' 
near the singularity\ $x^+\ra 0$.}.

We can also examine the behaviour of the zero modes or center-of-mass 
modes. For $n=0$, the equations of motion (\ref{EOM2}) for $X^I_0(\tau)$ 
can be solved to give
\be\label{XI0}
X^i_0(\tau)={x^i_0\over\sqrt{2\pi}} + \sqrt{2\pi} \al' p_{i0} \tau\ , \qquad 
X^m_0(\tau)={x^m_0\over\sqrt{2\pi}} + \sqrt{2\pi} \al' p_{m0} {\tau^{2\nu}}\ ,
\ee
where $p_{I0}$ are the center-of-mass momenta defined later (\ref{cmmomenta}).
These show that for singularities with $2\nu\geq 0$, the center of mass of the
string is not driven to infinity by the singularity. 
We will find parallels of this with the asymptotics of low lying string 
oscillation modes. This is to be contrasted with the divergences for 
spacetimes with $2\nu<0$.\ 
Note that the zero mode behaviour is essentially point-particle-like.
Thus the centers-of-mass of say a collection of infalling strings would 
appear to exhibit diverging tidal forces through geodesic deviation. 
However the crucial point is that the oscillations of the string are 
now non-negligible (even if finite). Thus neighbouring strings would 
appear to have large spatial overlap and string interactions become 
important near the singularity.

The mode expansion for the spacetime coordinates of the string is
\be
\label{modeexpXIn}
X^I(\tau,\sigma) = X^I_0(\tau) + \sum_{n=1}^\infty \left( 
k_n^I f^I_n(\tau) (a^I_n e^{in\sigma} + {\tilde a}^I_n e^{-in\sigma}) + 
k_n^{I*} f^{I*}_n(\tau) (a^I_{-n} e^{-in\sigma} + {\tilde a}^I_{-n} 
e^{in\sigma}) \right)\ .
\ee
The constant $k_n^I$ will be fixed by demanding canonical commutation 
relations for the creation-annihilation operators. The momentum 
conjugates\ $\Pi^I={\del {\cal L}\over\del ({\del_\tau X^I})}$\ are
\be\label{momenta}
\Pi^i(\tau,\sigma) = {1\over 2\pi\al'} {\del_\tau X}^i\ , \qquad
\Pi^m(\tau,\sigma) = {\tau^{b-a}\over 2\pi\al'} {\del_\tau X}^m\ .
\ee
We define the center-of-mass momenta $p_{I0}$ as
\be\label{cmmomenta}
p_{i0} = \int_0^{2\pi} {d\sigma\over\sqrt{2\pi}} \Pi^i 
= {1\over\sqrt{2\pi} \al'} {\dot X^i_0}(\tau)\ ,
\qquad
p_{m0} = \int_0^{2\pi} {d\sigma\over\sqrt{2\pi}} \Pi^m 
= {\tau^{b-a}\over\sqrt{2\pi} \al'} {\dot X^m_0}(\tau)\ .
\ee
Then we see that imposing the nonzero commutation relations 
\be\label{acommrelns}
[x^I_0, p_{J0}] = i\delta^I_J\ , \qquad 
[a^I_n,a^J_{-m}]=n\delta^{IJ}\delta_{nm}\ , \qquad 
[{\tilde a}^I_n,{\tilde a}^J_{-m}]=n\delta^{IJ}\delta_{nm}\ ,
\ee
implies the equal time commutation relations, \eg\
\be
[X^I(\tau,\sigma),\Pi^J(\tau,\sigma')] = {i\over 2\pi} \delta^{IJ} 
\left(1+\sum_{n=1}^\infty (e^{in(\sigma-\sigma')}+e^{-in(\sigma-\sigma')})
\right) = i \delta^{IJ} \delta(\sigma-\sigma')\ ,
\ee
using the Fourier series representation for the Dirac $\delta$-function, 
with the constant $k_n^I$ being (this agrees with the conventions of 
\cite{joetext} for flat space, except for a reversal of left/right movers)
\be
k_n^I = {i\over n}\ \sqrt{{\pi\al'\over 2|c^I_{n0}| (a+1)}}\ ,\qquad
c^I_{n0}=c^I_{n1}c^{I*}_{n2} - c^{I*}_{n1} c^I_{n2}\ ,
\ee
where $c^I_{n0}$ is the Wronskian.
We have used above the expressions for the derivatives of the mode 
functions $f^I_n$ and some recursion relations for the Bessel 
functions\footnote{We have used the following, the Bessel function 
argument being\ $({n\tau^{a+1}\over a+1})$,
\bea
&& {df^i_n(\tau)\over d\tau} =  n \sqrt{n} \tau^{a+{1\over 2}} 
\left( c^i_{n1} J_{{1\over 2a+2}-1} + c^i_{n2} Y_{{1\over 2a+2}-1} \right) ,
\qquad \frac{df^m_n(\tau)}{d\tau} =  n \sqrt{n} \tau^{a+\nu} 
\left( c^m_{n1} J_{{\nu\over a+1}-1} + c^m_{n2} Y_{{\nu\over a+1}-1} 
\right)\ ,\nonumber\\
&& J_{\nu-1}(z) + J_{\nu+1}(z) = {2\nu\over z} J_{\nu}(z)\ , \qquad
J_{\nu-1}(z) - J_{\nu+1}(z) = 2 {dJ_{\nu}(z)\over dz}\ , \nonumber \\
&& Y_{\nu-1}(z) + Y_{\nu+1}(z) = {2\nu\over z} Y_{\nu}(z)\ , \qquad
Y_{\nu-1}(z) - Y_{\nu+1}(z) = 2 {dY_{\nu}(z)\over dz} \ , \qquad
J_{\nu}(z) Y_{\nu-1}(z) - J_{\nu-1}(z) Y_{\nu}(z) = {2\over\pi z} \ .\nonumber
\eea
} to calculate the Wronskian of $f^I_n,{\dot f}^I_n$.

Let us now discuss level matching. The operator that generates 
$\sigma$-translations is worldsheet momentum $P$ given by the stress tensor
\be
T_{ab} \sim -{1\over \sqrt{-h}}\ {\delta{\cal L}\over\delta h^{ab}} 
\sim -\left( g_{IJ} \partial_aX^I \partial_bX^J 
- {1\over 2} h_{ab}\ h^{cd}  g_{IJ} \partial_cX^I \partial_dX^J \right)\ .
\ee
Then the $\sigma$-translation gauge invariance is fixed by demanding 
that the momentum operator vanishes on the physical states, \ie\ \
$P=\int d\sigma T_{\tau\sigma}=0$. From our action above and our lightcone 
gauge condition $h_{\tau\sigma}=0$, we have
\be
P=\int d\sigma (\tau^a \del_\tau X^i \del_\sigma X^i + 
\tau^b \del_\tau X^m \del_\sigma X^m)\ .
\ee
Using the mode expansion (\ref{modeexpXIn}), this can be evaluated as
\be\label{levelmatch}
P \sim \tau^a \sum_n n \Big( (a^i_{-n}a^i_n-{\tilde a}^i_{-n} {\tilde a}^i_n) 
+ (a^m_{-n}a^m_n-{\tilde a}^m_{-n} {\tilde a}^m_n) \Big)\ ,
\ee
where we have used the Bessel recursion relations and the expressions for 
${\dot f}^I_n$ (suppressing some overall unimportant numerical factors). 
This recovers the level matching conditions $N={\tilde N}$.

Now we calculate the string Hamiltonian. Using the mode expansion 
(\ref{modeexpXIn}), we first evaluate
\bea\label{intermediateterms}
{1\over 2\pi} \int_0^{2\pi} d\sigma (\del_\tau X^I)^2 &=& ({\dot X}^I_0)^2 
+ \sum_n  |k_n|^2 \Big( |{\dot f}^I_n|^2 ( \{a^I_n,a^I_{-n}\} + 
\{{\tilde a}^I_n,{\tilde a}^I_{-n}\} ) 
- ({\dot f}^I_n)^2 \{a^I_n,{\tilde a}^I_n\} \nonumber\\
&& {} \qquad\qquad\qquad\ 
- ({\dot f}^{I*}_n)^2 \{a^I_{-n},{\tilde a}^I_{-n}\} \Big)\ , \nonumber\\
{1\over 2\pi} \int_0^{2\pi} d\sigma (\del_\sigma X^I)^2 &=&
\sum_n n^2 |k_n|^2 \Big( |f^I_n|^2 ( \{a^I_n,a^I_{-n}\} + 
\{{\tilde a}^I_n,{\tilde a}^I_{-n}\} ) 
- (f^I_n)^2 \{a^I_n,{\tilde a}^I_n\}  \nonumber\\
&& {} \qquad\qquad\qquad\qquad\qquad
- (f^{I*}_n)^2 \{a^I_{-n},{\tilde a}^I_{-n}\} \Big)\ .
\eea
The Hamiltonian (\ref{Hamiltonian}) then simplifies to
\bea\label{Hamil}
H &=& {1\over 2\al'} \left(({\dot X}^i_0)^2 + \tau^{b-a}({\dot X}^m_0)^2\right)
\nonumber\\
&& +\ {1\over 2\al'} \sum_n |k_n|^2 \Biggl( ( \{a^i_n,a^i_{-n}\} + 
\{{\tilde a}^i_n,{\tilde a}^i_{-n}\} ) \left( |{\dot f}^i_n|^2 + n^2 \tau^{2a} 
|f^i_n|^2 \right) \nonumber\\
&& -\ \{a^i_n,{\tilde a}^i_n\} \left( ({\dot f}^i_n)^2 + n^2 \tau^{2a} 
(f^i_n)^2 \right) - \{a^i_{-n},{\tilde a}^i_{-n}\} 
\left( ({\dot f}^{i*}_n)^2 + n^2 \tau^{2a} (f^{i*}_n)^2 \right) \Biggr) 
\nonumber\\
&& +\ {1\over 2\al'} \sum_n |k_n|^2 \Biggl( ( \{a^m_n,a^m_{-n}\} + 
\{{\tilde a}^m_n,{\tilde a}^m_{-n}\} ) \left( \tau^{b-a} |{\dot f}^m_n|^2 
+ n^2 \tau^{b+a} |f^m_n|^2 \right) \nonumber\\
&& \qquad -\ \{a^m_n,{\tilde a}^m_n\} \left( \tau^{b-a} ({\dot f}^m_n)^2 
+ n^2 \tau^{b+a} (f^m_n)^2 \right) \nonumber\\
&& \qquad -\ \{a^m_{-n},{\tilde a}^m_{-n}\} 
\left( \tau^{b-a} ({\dot f}^{m*}_n)^2 + n^2 \tau^{b+a} (f^{m*}_n)^2 \right) 
\Biggr)\ .
\eea
In the next section, we will examine free string behaviour in the 
vicinity of the singularity.

\subsection{Strings in the near singularity region}

Let us now understand the behaviour of the string mode functions near 
the singularity. It turns out that the near singularity limit $\tau\ra 0$ 
must be taken with care. We define a cutoff $\tau=\tau_\epsilon\sim 0$ as 
a short time regulator in the vicinity of the singularity $\tau=0$. Then 
define\ $n_\epsilon\equiv {1\over \tau_\epsilon^{a+1}}$\ as a cutoff on the 
worldsheet oscillation number. We then see sharp differences between 
the behaviour near $\tau=\tau_\epsilon$ of string modes with ``low 
lying'' oscillation numbers\ $n\lesssim n_\epsilon$\ (\ie\ 
$n\tau_\epsilon^{a+1} \ll 1$), and highly oscillating string modes with 
$n\gg n_\epsilon$\ (\ie\ $n\tau_\epsilon^{a+1} \gg 1$).

Noting the asymptotics $J_{\pm \nu}(x) \sim x^{\pm \nu}$ 
for $x\sim 0$, and\ $Y_{\nu}=\cot(\pi\nu)J_\nu-{\rm cosec}(\pi\nu)J_{-\nu}$, 
we see that, near $\tau=0$, the $f^I_n(\tau)$ approach,
\be\label{asympfIn}
f^i_n \ra {\lambda^i_{n0}} + {\lambda^i_{n\tau}} \tau\ , \qquad 
f^m_n \ra {\lambda^m_{n0}} + {\lambda^m_{n\tau}} \tau^{2\nu}\ 
\qquad (\tau\ra 0)\ ,
\ee
for modes with low lying oscillation numbers\ $n\lesssim n_\epsilon$.
The constant coefficients are (from the asymptotic Bessel expressions)
\bea\label{c12lambda}
{\lambda^i_{n\tau}} = \sqrt{n} \left({n\over 2a+2}\right)^{{1\over 2a+2}}\ 
{{c^i_{n1}}+{c^i_{n2}}\cot {\pi\over 2a+2} \over \Gamma({2a+3\over 2a+2})}\ ,
\qquad
{\lambda^i_{n0}} 
= -{c^i_{n2}} \sqrt{n} \left({n\over 2a+2}\right)^{-{1\over 2a+2}}\ 
{{\rm cosec} {\pi\over 2a+2}\over \Gamma({2a+1\over 2a+2})}\ ,\nonumber\\
{\lambda^m_{n\tau}} = \sqrt{n} \left({n\over 2a+2}\right)^{{\nu\over a+1}}\ 
{{c^m_{n1}}+{c^m_{n2}}\cot {\nu\pi\over a+1}\over\Gamma({a+\nu+1\over a+1})}\ ,
\qquad
{\lambda^m_{n0}} 
= -{c^m_{n2}} \sqrt{n} \left({n\over 2a+2}\right)^{-{\nu\over a+1}}\ 
{{\rm cosec} {\nu\pi\over a+1}\over \Gamma({a+1-\nu\over a+1})}\ .\quad
\eea
Thus we see that the asymptotic $\tau$-dependence of such finite $n$
string oscillation modes near $\tau\ra 0$ is essentially the same as
for the the center-of-mass modes of the string (\ref{XI0}). Thus the
(classical) string mode amplitudes are non-divergent near the singularity 
for cosmological solutions with $2\nu=a+1-b\geq 0$. The string 
oscillation amplitude in such a curved spacetime is perhaps better 
defined as\ $g_{mm} (f^m_n)^2$:\ this gives the asymptotics to be 
non-divergent for $2a+2\geq b$.
In what follows, we will find useful the Wronskian combinations for the 
$\lambda^I_{n0},\lambda^I_{n\tau}$,
\bea\label{WronLambda}
&& \Lambda^i_{n,0\tau}\equiv
\lambda^i_{n0} \lambda^{i*}_{n\tau} - \lambda^i_{n\tau} \lambda^{i*}_{n0}
= n c^i_{n0}\ { {\rm cosec} {\pi\over 2a+2}\over \Gamma({2a+3\over 2a+2}) 
\Gamma({2a+1\over 2a+2})}\ ,\nonumber\\
&& \Lambda^m_{n,0\tau}\equiv
\lambda^m_{n0} \lambda^{m*}_{n\tau} - \lambda^m_{n\tau} \lambda^{m*}_{n0}
= n c^m_{n0}\ { {\rm cosec} {\nu\pi\over a+1}\over \Gamma({a+1+\nu\over a+1}) 
\Gamma({a+1-\nu\over a+1})}\ ,
\qquad c^I_{n0}=c^I_{n1}c^{I*}_{n2}-c^{I*}_{n1}c^I_{n2}\ .
\eea

On the other hand, consider now modes with $n\gg n_\epsilon$. Then we 
can see from the Bessel mode functions (\ref{modesolns})\ (or directly 
from the equations of motion (\ref{EOM2})) that these are oscillatory 
near the singularity:\ the argument ${n\tau^{a+1}\over a+1}$ cannot be 
taken to be small and the asymptotics (\ref{asympfIn}) above are not valid.\ 
For instance, choosing linear combinations $c^I_{n1},c^I_{n2}=1,\pm i$, 
gives modes that are the analogs of ingoing or outgoing plane waves, 
\ie\ the $f^I_n$ are Hankel functions dressed with powers of $\tau$, 
with asymptotics\footnote{This is also the asymptotic behaviour near 
$\tau\ra \infty$ of the modes (\ref{modesolns}) for $any$ $n$.} for 
$\tau\ra 0$
\be\label{nlargefIn}
f^i_n \sim {1\over \tau^{a/2}}~e^{\pm in\tau^{a+1}/(a+1)}\ , \qquad 
f^m_n \sim  {1\over \tau^{b/2}}~e^{\pm in\tau^{a+1}/(a+1)}\ , 
\qquad\  n\gg n_\epsilon\ .
\ee
Note that for any regulator $\tau_\epsilon$, however small, in the 
vicinity of the singularity, there exist modes of sufficiently high 
oscillation $n$ such that the corresponding modes $f^I_n$ are of this 
form (\ref{nlargefIn}). Since the string oscillation number $n$ can be 
arbitrarily large, such modes exist uniformly for all singularities, 
with $2\nu\gtrless 0$, and are in a sense transplanckian:\ they are 
reminiscent of high frequency scalar modes propagating in an 
inflationary background.
This behaviour, somewhat different from the finite $n$ mode behaviour, 
is distinctly stringy.

We first analyse the case\ $2\nu=a+1-b\geq 0$.\ 
Using the asymptotic forms of the mode functions $f^I_n$ near the 
singularity $\tau\ra 0$, for finite $n\lesssim n_\epsilon$ modes,
\be
f^i_n\ra {\lambda^i_{n0}}\ , \qquad {\dot f}^i_n\ra {\lambda^i_{n\tau}}\ , 
\qquad 
f^m_n\ra {\lambda^m_{n0}}\ , 
\qquad {\dot f}^m_n\ra {\lambda^m_{n\tau}}\ (2\nu)\ \tau^{2\nu-1}\ ,
\ee
the Hamiltonian (\ref{Hamil}) simplifies to
\bea\label{Hamil2}
H &=& {1\over 2\al'} \left(({\dot X}^i_0)^2 + \tau^{b-a}({\dot X}^m_0)^2\right)
\nonumber\\
&& +\ {1\over 2\al'} \sum_n |k_n|^2 \Biggl( \left( ( \{a^i_n,a^i_{-n}\} + 
\{{\tilde a}^i_n,{\tilde a}^i_{-n}\} ) |{\lambda^i_{n\tau}}|^2 
-\ \{a^i_n,{\tilde a}^i_n\} ({\lambda^i_{n\tau}})^2 
- \{a^i_{-n},{\tilde a}^i_{-n}\} ({\lambda^{i*}_{n\tau}})^2 \right) 
\nonumber\\
&& \quad +\ n^2 \tau^{2a} \left( ( \{a^i_n,a^i_{-n}\} + 
\{{\tilde a}^i_n,{\tilde a}^i_{-n}\} ) |{\lambda^i_{n0}}|^2 
-\ \{a^i_n,{\tilde a}^i_n\} ({\lambda^i_{n0}})^2 
- \{a^i_{-n},{\tilde a}^i_{-n}\} ({\lambda^{i*}_{n0}})^2 \right) \Biggr)
\nonumber\\
&& +\ \sum_n {|k_n|^2\over 2\al'} \Biggl( \tau^{a-b} (2\nu)^2 
\left( ( \{a^m_n,a^m_{-n}\} + 
\{{\tilde a}^m_n,{\tilde a}^m_{-n}\} ) |{\lambda^m_{n\tau}}|^2 
- \{a^m_n,{\tilde a}^m_n\} ({\lambda^m_{n\tau}})^2 
- \{a^m_{-n},{\tilde a}^m_{-n}\} ({\lambda^{m*}_{n\tau}})^2 \right) 
\nonumber\\
&& \quad +\ n^2 \tau^{b+a} \left( ( \{a^m_n,a^m_{-n}\} + 
\{{\tilde a}^m_n,{\tilde a}^m_{-n}\} ) |{\lambda^m_{n0}}|^2 
-\ \{a^m_n,{\tilde a}^m_n\} ({\lambda^m_{n0}})^2 
- \{a^m_{-n},{\tilde a}^m_{-n}\} ({\lambda^{m*}_{n0}})^2 \right) \Biggr)\ .
\eea
Note that there are ``interaction terms'' of the form\ 
$a^I_n {\tilde a}^I_n$\ and\ $a^{I\dag}_n {\tilde a}^{I\dag}_n$\ besides 
the diagonal number-operator terms. The interaction terms have the same 
$\tau$-dependent coefficients as the diagonal terms so that they are 
not unimportant and cannot be ignored.

The corresponding calculation for flat space ($a,b=0$) involves sine
and cosine modes (the analogs of the Bessel-$J,Y$), the Hamiltonian
having no time-dependence. Analysing this near $\tau\ra 0$, we see 
that the coefficients of the $a^I_n{\tilde a}^I_n$- and 
$a^I_{-n}{\tilde a}^I_{-n}$-terms are of the form\ $(c_1^2+c_2^2)$
and $(c_1^{*2}+c_2^{*2})$, while that of the diagonal terms is 
$(|c_1|^2+|c_2|^2)$, where $c_1,c_2$ are the coefficients of the 
sine, cosine: then we see that choosing the usual positive frequency 
modes with $c_1,c_2$ being $1,-i$, results in just the diagonal term 
in the Hamiltonian. In the present case, due to the extra 
$\tau$-dependences in the Hamiltonian, the resulting expressions do 
not simplify and the ``interaction'' terms remain. A similar 
calculation with different choices of the basis modes (\eg\ Hankel 
functions) yields equivalent results.

This Hamiltonian (\ref{Hamil2}), corresponding to the choice of $x^+$ as 
a time coordinate\footnote{The affine parameter quantization, Appendix B,
yields similar results as we describe here.}, can now be recast as
\bea\label{Hamil1}
H &=& \pi\al' ((p_{i0})^2 + \tau^{a-b}(p_{m0})^2)
+\ \sum_n {\pi\over 2(a+1) n^2}\Biggl( {1\over |c^i_{n0}|} \left( 
b^{i\dag}_{n\tau} b^i_{n\tau} +\ n^2 \tau^{2a} b^{i\dag}_{n0} b^i_{n0} \right)
 \nonumber\\
&& \qquad\qquad\qquad\qquad\qquad\qquad
+\ {1\over |c^m_{n0}|} \left((2\nu)^2\tau^{a-b} b^{m\dag}_{n\tau} b^m_{n\tau} 
+\ n^2 \tau^{b+a} b^{m\dag}_{n0} b^m_{n0} \right) \Biggr)\ ,
\eea
where we have defined new oscillator modes (and their Hermitian conjugates)
\be\label{bmodes}
b^I_{n0} = \lambda^I_{n0} a^I_n - \lambda^{I*}_{n0}\ {\tilde a}^I_{-n}\ , 
\qquad  b^I_{n\tau} 
= \lambda^I_{n\tau} a^I_n - \lambda^{I*}_{n\tau}\ {\tilde a}^I_{-n}\ ,
\qquad I=i,m\ .
\ee
The string oscillator masses are Lorentz invariant expressions 
\be\label{mass}
m^2 = - 2 g^{+-} p_+ p_- - g^{II} (p_{I0})^2\ .
\ee
From the above expressions, and recalling that $p_-=-{1\over 2\pi\al'} ,\
-p_+=H$, we see that the center-of-mass terms cancel resulting in the 
time-dependent masses for these low-lying $n\lesssim n_\epsilon$ 
oscillation string modes
\be\label{masses}
m^2(\tau) = {1\over 2\al'(a+1)}\ \sum_{i,m;\ n\lesssim n_\epsilon} 
\left( {1\over\tau^a} {N^i_{n\tau} \over n^2 |c^i_{n0}|} 
+ \tau^a {N^i_{n0} \over |c^i_{n0}|} + {(2\nu)^2\over\tau^b} 
{N^m_{n\tau}\over n^2 |c^m_{n0}|} + \tau^b {N^m_{n0}\over |c^m_{n0}|} 
\right)\ , \qquad\quad [2\nu\geq 0]\ ,
\ee
defining
\be
N^i_{n\tau} = b^{i\dag}_{n\tau} b^i_{n\tau}\ ,\quad
N^i_{n0} = b^{i\dag}_{n0} b^i_{n0}\ , \qquad
N^m_{n\tau} = b^{m\dag}_{n\tau} b^m_{n\tau}\ ,
\quad N^m_{n0} = b^{m\dag}_{n0} b^m_{n0}\ .
\ee
These expressions should be understood as valid in the vicinity of 
the singularity, but only upto the regulator\ $\tau\lesssim \tau_\epsilon$.\\
The original left- and right-moving oscillator operators can be re-expressed
in terms of $b^I_n$ as 
\be
a^I_n = {1\over \Lambda^I_{n,0\tau}}\ \left( \lambda^{I*}_{n\tau}\ b^I_{n0} 
- \lambda^{I*}_{n0}\ b^I_{n\tau} \right)\ , \qquad
{\tilde a}^I_n = {1\over \Lambda^{I*}_{n,0\tau}}\ 
\left( \lambda^{I*}_{n\tau}\ b^{I\dag}_{n0} - 
\lambda^{I*}_{n0}\ b^{I\dag}_{n\tau} \right)\ ,\qquad I=i,m\ ,
\ee
and the level matching condition (\ref{levelmatch}) is recast as
\be\label{levelmatchb}
0 = \sum_n n (a^I_{-n} a^I_n - {\tilde a}^I_{-n} {\tilde a}^I_n)
= \sum_n {n\over \Lambda^I_{n,0\tau}}\ 
(b^{I\dag}_{n\tau} b^I_{n0} - b^I_{n\tau} b^{I\dag}_{n0})\ .
\ee
The commutation relations satisfied by the $b^I_{n0} , b^I_{n\tau}$ are
\bea\label{bcomm}
[b^I_{m0},b^{J\dag}_{n0}] = 0 = [b^I_{m\tau},b^{J\dag}_{n\tau}] = 
[b^I_{m0},b^J_{n\tau}]\ , && \quad 
[b^I_{m0},b^{J\dag}_{n\tau}] = n\Lambda^I_{n,0\tau}\delta^{IJ}\delta_{mn} = 
-[b^I_{n\tau},b^{J\dag}_{n0}]\ ,\nonumber\\
{} [N^I_{m0},b^J_{n\tau}] = n \Lambda^I_{n,0\tau} b^I_{n0} 
\delta^{IJ}\delta_{mn} \ , && \quad 
[N^I_{m0},b^{J\dag}_{n\tau}] = n \Lambda^I_{n,0\tau} b^{I\dag}_{n0}
\delta^{IJ}\delta_{mn} \ ,\nonumber\\
{} [N^I_{m\tau},b^J_{n0}] = -n \Lambda^I_{n,0\tau} b^I_{n\tau} 
\delta^{IJ}\delta_{mn} \ ,  && \quad 
[N^I_{n\tau},b^{I\dag}_{n0}] = -n \Lambda^I_{n,0\tau} b^{I\dag}_{n\tau}\ ,
\nonumber\\
{} [N^I_{m0},N^J_{n\tau}] &=& n\delta^{IJ}\delta_{mn} 
\Lambda^I_{n,0\tau} (b^{I\dag}_{n0}b^I_{n\tau}+b^{I\dag}_{n\tau}b^I_{n0})\ .
\eea
using the left- and right-moving $a,{\tilde a}$-oscillator commutators 
(\ref{acommrelns}), and the Wronskian combinations $\Lambda^I_{n,0\tau}$s 
from (\ref{WronLambda}).
Since the $b^I_{n0}, b^I_{n\tau}$, operators commute with their conjugates, 
the $N^I_{n0},N^I_{n\tau}$ operators do not have a number-operator-like 
interpretation on states annihilated by $b^I_{n0}, b^I_{n\tau}$.\\
From the expression for the time-dependent masses, it is tempting 
to speculate that states that have \eg\ vanishing 
$\langle N^i_{n\tau} \rangle$ but nonzero $\langle N^i_{n0}\rangle$ 
will become massless near the singularity $\tau\ra 0$. However since 
$N^I_{n\tau},N^I_{n0}$, do not commute\footnote{In terms of the original 
$a,{\tilde a}$-operators, this expression is
\bea
[N^I_{n0},N^J_{n\tau}]=n\delta^{IJ} \big[(\lambda^I_{n0}\lambda^{I*}_{n\tau} 
+ \lambda^I_{n\tau}\lambda^{I*}_{n0}) (a^I_{-n} a^I_n + 
{\tilde a}^I_n {\tilde a}^I_{-n}) - 2\lambda^I_{n0}\lambda^I_{n\tau} a^I_n
{\tilde a}^I_n - 2\lambda^{I*}_{n0}\lambda^{I*}_{n\tau} a^I_{-n}
{\tilde a}^I_{-n}\big] .\nonumber
\eea
}, these are generically not simultaneous eigenstates of $N^I_{n0}$ 
and $N^I_{n\tau}$, or energy eigenstates. 
If such a possibility can be validated for these $b^I$-states, 
then the $b^i_{n0},b^m_{n\tau}$-oscillator states are light near the 
singularity while the $b^i_{n\tau},b^m_{n0}$-oscillator states are 
massive near $\tau=x^+\ra 0$, for the singularities with $2\nu\geq 0, 
b<0$\ (while for $b>0$ singularities, the $b^I_{n0}$-oscillator 
states are light and the $b^I_{n\tau}$-states are massive). 
All these are light relative to the typical curvature scale however, 
as we will outline later. Some description of the $b^I$-states is 
given in the next subsection: it would be interesting to develop 
ths further.

Let us now consider the case\ $2\nu=a+1-b<0$. Then the modes $f^m_n$ 
behave near $\tau\ra 0$ as\ $f^m_n\ra {\lambda^m_{n\tau}}\ \tau^{2\nu}$ ,
while ${\dot f^m_n}\ra {\lambda^m_{n\tau}}\ (2\nu)\ \tau^{2\nu}$. Thus 
$\lambda^m_{n0}$ does not appear in the Hamiltonian (\ref{Hamil}) 
evaluated near $\tau\ra 0$, which thus shows all $a^m, 
{\tilde a^m}$-terms having identical asymptotics with time-dependence 
as $\tau\ra 0$ \eg
\be
\tau^{b-a} |{\dot f}^m_n|^2 + n^2 \tau^{b+a} |f^m_n|^2\ \sim\ 
\tau^{a-b} ((2\nu)^2 + n^2 \tau^{2a+2})\ \ra\ \tau^{a-b}
\ee
in the coefficients. It is therefore not particularly insightful to
recast $a^m,{\tilde a^m}$ in terms of the $b^m$-operators. The invariant 
oscillator masses thus grow as\ ${1\over\tau^a}$ and ${1\over\tau^b}$ 
for the $b^i_{n\tau}$- and $a^m$-oscillator states. The $b^I_{n0}$-states 
are light as before.

Now let us consider the high oscillation modes with\ 
$n\gg n_\epsilon={1\over\tau_\epsilon^{a+1}}$:\ these have a uniform 
behaviour for both $2\nu\gtrless 0$. Then using the asymptotics 
(\ref{nlargefIn}) for such modes (with $c^I_{n1}=1, c^I_{n2}=-i$, which 
are positive frequency), we see that
\be\label{fdotlargen}
{\dot f}^i_n \sim\ \left( -in\tau^a-{a\over 2\tau} \right) 
{e^{-in\tau^{a+1}/(a+1)}\over\tau^{a/2}}\ , \qquad
{\dot f}^m_n \sim\ \left( -in\tau^a-{b\over 2\tau} \right) 
{e^{-in\tau^{a+1}/(a+1)}\over \tau^{b/2}}\ .
\ee
This is very similar to the asymptotics of the modes (\ref{modesolns}) 
at early times\ $|\tau|\ra\infty$: however we are considering a different 
limit here, with\ large $n$, small $\tau$, and\ $n\tau^{a+1}\gg 1$, 
so it is worth elaborating a little. 
In this limit, we calculate the expressions in (\ref{Hamil}) and 
express them as
\bea
{1\over n^2} \left(({\dot f}^i_n)^2 + n^2 \tau^{2a} (f^i_n)^2\right) \sim\
\tau^a \left( {a^2\over 4(n\tau^{a+1})^2} + {ia\over(n\tau^{a+1})} \right)
e^{-2in\tau^{a+1}/(a+1)}\ , \nonumber\\
{1\over n^2} \left(\tau^{b-a} ({\dot f}^m_n)^2 + n^2 \tau^{b+a} (f^m_n)^2
\right) \sim\ \tau^a \left( {b^2\over 4(n\tau^{a+1})^2} 
+ {ib\over(n\tau^{a+1})} \right) e^{-2in\tau^{a+1}/(a+1)}\ , \nonumber\\
{1\over n^2} \left(|{\dot f}^i_n|^2 + n^2 \tau^{2a} |f^i_n|^2\right)
\sim\ 2\tau^a\ , \qquad
{1\over n^2} \left(\tau^{b-a}|{\dot f}^m_n|^2 + n^2\tau^{b+a}|f^m_n|^2\right)
\sim\ 2\tau^a\ .
\eea
The expressions in the first two lines are vanishingly small relative 
to the ones in the third, and the Hamiltonian (\ref{Hamil}) simplifies to
\be\label{Hlargen}
H_{n\gg n_\epsilon} \sim\ \tau^a \sum_{I;\ n\gg n_\epsilon}\ {1\over a+1}\ 
(a^I_{-n} a^I_n + {\tilde a}^I_{-n} {\tilde a}^I_n + n)\ ,
\ee
as for free string propagation. The overall factor $\tau^a$ arises as 
before from the fact that we are using $x^+$ as time coordinate (with 
$g_{+-}=-(x^+)^a$). The oscillator masses for these highly stringy modes, 
using (\ref{mass}), become
\be
m^2(\tau) \sim\ -g^{+-} H {1\over\al'} 
= \sum_{I;\ n\gg n_\epsilon}\ {1\over a+1}\ 
(N^I_n + {\tilde N}^I_n + n)\ ,
\ee
as for free strings in flat space. The zero point energy has an 
ultraviolet completion as in that case. Thus these highly stringy modes
exhibit essentially free propagation in these backgrounds. Comparing 
the  mass\ $\sqrt{{n\over\al'}}$\ of a typical single excitation state
with the typical curvature scale set by the tidal forces $|a^i|,|a^m|$, 
in this region, we have\ 
${n\over\al' |a^i|^2} \sim\ {n\tau^{3a+4}\over\al'} ,\
{n\over\al' |a^m|^2} \sim\ {n\tau^{4a+4-b}\over\al'}$.\ Thus states 
satisfying\ ${1\over\tau^{a+1}}\ll n\ll {1\over\tau^{3a+4}}$\ and\ 
${1\over\tau^{a+1}}\ll n\ll {1\over\tau^{4a+4-b}}$\ are light relative 
to the local curvature scale. Similar comparisons for the 
$n\lesssim n_\epsilon$ states with the $\tau$-dependences\ 
$\tau^{\pm a}$ and $\tau^{\pm b}$ relative to the typical curvature 
scale hold if $b<2a+2$.

For any finite, if infinitesimal, value of the near-singularity
cutoff\ $\tau_\epsilon$, such highly stringy modes exist, for
oscillation number\ $n\gg n_\epsilon={1\over\tau_\epsilon^{a+1}}$ , 
although naively removing the cutoff would suggest the absence of any 
such modes.

\subsection{Near singularity string states and wavefunctions}

We describe here some aspects of string states near the singularity
using our discussion in the previous section, beginning with the 
low lying oscillation mode $b^I$-states.

Noting that the $b^I$ operators are complex linear combinations of the
$a^I,{\tilde a}^{I\dag}$, and recalling Bogolubov transformations, a
$b^I$-vacuum $|\phi\rangle$ (annihilated by the $b^I$) would appear to
be a multi-particle state in terms of the original $a^I,{\tilde a}^I$
operators, and vice-versa. Indeed we have\
\be
\langle 0| \sum_n N^I_{n\rho} |0\rangle = \sum_n n |\lambda^I_{n\rho}|^2\ ,
\qquad\qquad \rho=0,\tau\ ,
\ee
where $a^I_n|0\rangle=0={\tilde a}^I_n|0\rangle$.\ Similarly, defining\ 
$|b_0\rangle=b^i_{n0}|0\rangle ,\ |b^\dag_0\rangle=b^{i\dag}_{n0}|0\rangle ,\ 
|b_\tau\rangle=b^i_{n\tau}|0\rangle ,\ 
|b^\dag_\tau\rangle=b^{i\dag}_{n\tau}|0\rangle$, it is straightforward to 
show that the lowest excited states have
\be
\langle b_p|b^{i\dag}_{mq}b^i_{mq}|b^\dag_r\rangle=0\ ,\quad
\langle b_p|b^{i\dag}_{mq}b^i_{mq}|b_r\rangle=n\lambda^{i*}_{nr}
\lambda^{i}_{np}\ ( \sum_m m|\lambda^i_{mq}|^2+n|\lambda^i_{nq}|^2 )\ ,
\quad p,q,r=0,\tau\ ,
\ee
using the expressions for the $b^I$ in terms of the $a^I,{\tilde a}^I$, 
and their commutation relations.

Now it can be shown that\ 
$[N^I_{m0},(b^{J\dag}_{n\tau})^l] = l (n\Lambda^{I*}_{n,0\tau}) \
\delta^{IJ} \delta_{nm}\ (b^{I\dag}_{n\tau})^{l-1}\ b^{I\dag}_{n0}$, using 
the $b^I$-oscillator algebra (\ref{bcomm}). Assuming the existence of 
a $b^I_0$-vacuum, defining an excited state\ 
$|\Psi_l\rangle = (b^{I\dag}_{n\tau})^l |\phi\rangle$\ gives\ 
$N^I_{n0}|\Psi_l\rangle = 
l (n\Lambda^{I*}_{n,0\tau}) b^{I\dag}_{n0}|\Psi_{l-1}\rangle$ .\\
Thus we see heuristically that, starting with the $b^I_0$-vacuum and 
constructing a Fock space using $b^{I\dag}_{n\tau}$, we obtain states 
with nonzero $\langle N^I_0\rangle$ . Similarly possible coherent 
states of the form\ $|s\rangle=e^{s b^{I\dag}_{n\tau}}|\phi\rangle$\ 
have\ $b_0|s\rangle \sim s|s\rangle$, upto numerical factors. 
Since $[b^I_0,b^{J\dag}_0]=0$, we see that\ $b^{I\dag}_{n0}|s\rangle$\ 
is also a coherent state with the same eigenvalue.

Note that these are not eigenstates of the Hamiltonian\ 
$H_{n\lesssim n_\epsilon}$ since $N^I_0, N^I_\tau$, do not commute, so 
generically such states mix under time evolution. Consider the 
Schrodinger equation\ \ $i{d\over d\tau} |\Psi\rangle = H|\Psi\rangle $,\  
with $|\Psi\rangle=\sum_l c_l|\Psi^i_l\rangle$\ constructed using only 
$b^{i\dag}_{n\tau}$-oscillators. This gives\
$i{d\over d\tau} |\Psi\rangle \sim\ \sum_{i,m,n} (N^i_{n\tau} + 
\tau^{2a} N^i_{n0}) |\Psi\rangle$ .\
This suggests that the time-dependence of these states is regular near 
the singularity $\tau\ra 0$. \\
Along similar lines, we can, more simply, construct states of the form\
$(b^{i\dag}_0)^{l_i} (b^{m\dag}_0)^{l_m} |\phi\rangle$, starting with the 
$b^I_0$-vacuum. These states have vanishing $\langle N^I_0\rangle$ but 
nonzero $\langle N^I_\tau\rangle$. The Schrodinger equation for such 
states is of the form\ $i{d\over d\tau} |\Psi\rangle
\sim\ \sum_{i,m,n} (N^i_{n\tau}+\tau^{a-b}N^m_{n\tau})|\Psi\rangle$. In 
accord with level matching (\ref{levelmatchb}), we can construct
states of the form\ $(b^I_\tau)^l (b^{J\dag}_0)^m|\phi\rangle$: then 
since $b^I_0,b^J_\tau$ commute, these states again have vanishing 
$\langle N^I_0\rangle$ and nonzero $\langle N^I_\tau\rangle$.

We have described states constructed in terms of the $b^I_0$-vacuum so
far: similarly assuming formally the existence of a vacuum annihilated 
by $b^I_{n\tau}$, we can construct excited states along the lines of 
arguments similar to the ones above.

To obtain some rudimentary intuition for the spacetime description of 
these states, let us now describe position space wave-functions near 
the singularity. We will analyze the wave-functions for the reduced 
quantum mechanics of string modes with $\sigma$-momentum $n$,
\bea
x^I_n = i|k^I_n| (f^I_n(\tau) a^I_n - f^{I*}_n(\tau) a^I_{-n})\ , 
\qquad
{\tilde x}^I_n = i|k^I_n| (f^I_n(\tau) {\tilde a}^I_n - f^{I*}_n(\tau) 
{\tilde a}^I_{-n})\ ,  \nonumber\\
\Pi^i_n = {i|k^i_n|\over 2\pi\al'}\left({\dot f}^i_n(\tau) a^i_n 
- {\dot f}^{i*}_n(\tau) a^i_{-n} \right)\ ,\quad
\Pi^m_n = {i|k^m_n|\tau^{b-a}\over 2\pi\al'}\left({\dot f}^m_n(\tau) a^m_n 
- {\dot f}^{m*}_n(\tau) a^m_{-n} \right)\ ,
\eea
from the string coordinate mode expansion (\ref{modeexpXIn}) and the 
momentum conjugates (\ref{momenta})\ (we have suppressed explicitly 
writing the left-moving momenta ${\tilde \Pi}^I_n$).

Transforming to a position-space Schrodinger representation, we set\ 
$\Pi^I_n=-i\del_{x^I_n} ,\ {\tilde \Pi}^I_n=-i\del_{{\tilde x}^I_n}$. 
It is then straightforward to obtain the expressions
\be\label{aadag}
a^i_n = {{\dot f}^{i*}_n x^i_n - 2\pi\al' f^{i*}_n (-i\del_{x^i_n})
\over i|k^i_n| (f^i_n {\dot f}^{i*}_n - {\dot f}^i_n f^{i*}_n)}\ , \qquad
a^m_n = {{\dot f}^{m*}_n x^m_n - 2\pi\al' \tau^{a-b} f^{m*}_n (-i\del_{x^m_n})
\over i|k^m_n| (f^m_n {\dot f}^{m*}_n - {\dot f}^m_n f^{m*}_n)}\ , 
\ee
and their conjugates, with similar expressions for the ${\tilde a}^I_n$.
We can obtain expressions for the $b^I$-oscillators from the 
definitions (\ref{bmodes}), mixing the left- and right-moving terms.
\bea
b^i_{n0} = {\lambda^i_{n0} \lambda^{i*}_{n\tau} x^i_n - 
\lambda^{i*}_{n0} \lambda^i_{n\tau} {\tilde x}^i_n + i |\lambda^i_{n0}|^2 
2\pi\al' (\del_{x^i_n}-\del_{{\tilde x}^i_n})\over i|k^i_n| 
\Lambda^I_{n,0\tau}}\ , \nonumber\\
b^i_{n\tau} = {|\lambda^i_{n\tau}|^2 (x^i_n - {\tilde x}^i_n) 
+ i 2\pi\al' (\lambda^{i*}_{n0} \lambda^i_{n\tau} \del_{x^i_n} 
- \lambda^i_{n0} \lambda^{i*}_{n\tau} \del_{{\tilde x}^i_n})\over i|k^i_n| 
\Lambda^I_{n,0\tau}}\ .
\eea
Then the ground state wave-function $|0\rangle$ defined as\ 
$a^I_n|0\rangle = 0 ,\ {\tilde a}^I_n|0\rangle = 0$, for low-lying 
oscillation modes, satisfies near the singularity
\be
 (\lambda^{i*}_{n\tau} x^i_n - 2\pi\al'\lambda^{i*}_{n0} (-i\del_{x^i_n})
\psi^i_0(x^I_n) = 0 = (\lambda^{i*}_{n\tau} {\tilde x}^i_n - 
2\pi\al'\lambda^{i*}_{n0} (-i\del_{{\tilde x}^i_n}) \psi^i_0(x^I_n)\ ,
\ee
giving\ $\psi^i_0(x^I_n)\sim exp[i{\lambda^{i*}_{n\tau}\over 
2\pi\al'\lambda^{i*}_{n0}} ((x^i_n)^2+({\tilde x}^i_n)^2)]$. For positive 
frequency modes with\ $c^i_{n1}=1, c^i_{n2}=-i$, we see that this 
simplifies to a real Gaussian part (as expected for a set of harmonic 
oscillators) and a phase containing\ $\cos({\pi\over 2a+2}) 
{\Gamma({2a+1\over 2a+2})\over \Gamma({2a+3\over 2a+2})}$ (this phase 
vanishes for flat space $a=0$). Note that there is no explicit 
$\tau$-dependence here: the wavefunction is regular near the 
singularity $\tau\ra 0$. Similar statements hold for the $x^m$-part 
of the wavefunction (if $2\nu>0$).\\
Excited states can be constructed using either the 
$a,{\tilde a}$- or the $b^I$-oscillators: these generically mix, as 
can be seen either from the interaction terms in the Hamiltonian, or 
alternatively by noting that the $b^I$ do not commute.

The highly stringy states are simpler to describe: they are simply 
states of the form\ 
\be
|k^I_n,{\tilde k}^J_n\rangle \equiv\ \prod_{I,J;\ n\gg n_\epsilon} 
(a^I_{-n})^{k^I_n} ({\tilde a}^I_{-n})^{{\tilde k}^I_n} |0\rangle\ .
\ee
These are in fact eigenstates of the Hamiltonian $H_{n\gg n_\epsilon}$ 
so their time evolution is relatively simple, with the Schrodinger 
equation giving
\be
i{d\over d\tau} |k^I_n,{\tilde k}^J_n\rangle\ \sim\ 
\tau^a (nk^I_n + {\tilde k}^J_n + n) |k^I_n,{\tilde k}^J_n\rangle\ .
\ee
This can be recast as\ \ $i{d\over d\lambda} |\Psi\rangle = H_\lambda 
|\Psi\rangle$ in terms of the affine parameter (\ref{affineparam}), 
with the corresponding quantization discussed in Appendix B. This 
equation is essentially of the same form as in flat space, with the
time parameter being the affine parameter: the time evolution is
essentially given by phases of the form\ 
$e^{-iE_{(k^I_n,{\tilde k}^J_n)} {\tau^{a+1}\over a+1}}=
e^{-iE_{(k^I_n,{\tilde k}^J_n)} \lambda}$ .

For the position space description of the highly stringy states, we 
need to evaluate the expressions taking the limit in question carefully: 
using (\ref{fdotlargen}), (\ref{aadag}), the ground state 
$\psi^{n\gg n_\epsilon}_0(x^I_n)$ annihilated by $a^I_n,{\tilde a}^I_n$, 
satisfies 
\be
\left( (in\tau^a - {a\over 2\tau}) x^i_n - 2\pi\al' (-i\del_{x^i_n})\right) 
\psi^{n\gg n_\epsilon}_0 = 0 = \left( (in\tau^a - {a\over 2\tau}) 
{\tilde x}^i_n - 2\pi\al' (-i\del_{{\tilde x}^i_n})\right) 
\psi^{n\gg n_\epsilon}_0\ ,
\ee
giving
\be
\psi^{n\gg n_\epsilon} \sim\ exp~[-(n\tau^a+{ia\over 2\tau}) 
((x^i_n)^2+({\tilde x}^i_n)^2)] = exp~[-n\tau^a (1+{ia\over 2n\tau^{a+1}}) 
((x^i_n)^2+({\tilde x}^i_n)^2)]\ .
\ee
Note that the phase, containing ${1\over\tau}$ , oscillates ``wildly'' 
as $\tau\ra 0$. However from the second expression, we see that in the 
limit we are considering, $n\tau^{a+1}\gg 1$, the phase oscillation is 
slower than the damping of the real gaussian part of the wavefunction.

Similarly\ $a^m_n\psi_0(x)=0={\tilde a}^m_n\psi_0(x)$\ gives\ 
$\psi_0(x) \sim\ exp~[-n\tau^a \tau^{b-a} (1+{ib\over 2n\tau^{a+1}})\ 
{((x^m_n)^2+({\tilde x}^m_n)^2\over 2}]$. Thus the overall factor\ 
$n\tau^a \tau^{b-a}=n\tau^b$\ is heavily damped for $b<0$, while the 
phase of the wavefunction is\ $\tau^{-2\nu}$, but damped relative to 
its real gaussian part.

Excited states can be constructed by acting with the creation operators:\ 
\eg\ the first excited states are\ \eg\ $a^i_{-n} {\tilde a}^j_{-n}\psi_0(x)
\sim\ x^i_n {\tilde x}^i_n e^{-{2in\tau^{a+1}\over a+1}} \psi_0(x)$.

As we have mentioned in the previous subsection, the near singularity
limit we are considering appears subtle. In particular, as time
evolves towards the singularity and $\tau_\epsilon$ shrinks, the
worldsheet oscillation number cutoff $n_\epsilon$ increases and these
highly stringy states are no longer eigenstates, except for $n$ larger
than the increased value of the cutoff\
$n_\epsilon(\tau_\epsilon-\delta\tau_\epsilon)$.  A state with some
$n_0\gg n_\epsilon(\tau_\epsilon)$ at some later time crosses the
cutoff threshold and ceases to be highly stringy: it then becomes part
of the set of $b^I$-states and interacts nontrivially with them. Since
there is an infinity of highly stringy modes, it would appear that 
this process will continue indefinitely: making the description of
changing the cutoff more precise might draw parallels with the
renormalization group. It would be interesting to understand this
better.

\section{Discussion}

We have constructed cosmological spacetimes with null Kasner-like
singularities: the Kasner exponents satisfy algebraic conditions
following from the Einstein equations satisfied by the backgrounds.
These near singularity spacetimes can be extrapolated to approximate
solutions that are asymptotically flat at early times. It is possible
to recast these as anisotropic plane-wave spacetimes, with the
corresponding $\al'$-exactness properties of higher derivative
corrections.

We have found that the classical string modes admit exact solutions in
terms of Bessel functions. Using the near singularity behaviour of the
string mode functions, we can analyse the lightcone string worldsheet
spectrum through the Hamiltonian and calculate the oscillator masses. 
The near singularity region, regulated by say $\tau<\tau_\epsilon$, 
always contains highly stringy modes with oscillation number 
$n\gg {1\over\tau_\epsilon^{a+1}}$ that propagate essentially freely in 
the background. On the other hand, low lying string modes (finite 
$n\lesssim {1\over\tau_\epsilon^{a+1}}$) have asymptotic 
near-singularity $\tau$-dependence similar to the center-of-mass 
mode. The oscillator masses are time-dependent and can be recast 
in terms of two new sets of oscillators, one of which becomes light. 
It would be interesting to explore this further.\\
This suggests that the vicinity of the singularity is filled with 
``stringy fuzz'', comprising highly stringy modes. We expect string 
interactions are non-negligible near the singularity.

Our analysis is essentially from the bosonic parts of the string
worldsheet theory. Since the worldsheet fermion terms are quadratic
(with covariant derivatives) for these purely gravitational 
backgrounds, we expect that including them will not qualitatively 
change our results here. It would be interesting to carry out the 
superstring analysis in detail. Relatedly, several aspects of the 
matrix string analysis in these backgrounds have been studied in 
\cite{blauMpp}.

We have largely been studying the near singularity Kasner-like 
spacetimes. Consider the case where the spacetime scale factor $e^f\ra 1$ 
asymptotically (so that the spacetime is flat at early times). To 
elaborate, note that from the equation of motion $R_{++}=0$, there is 
a function-worth of solutions, \ie\ for a generic $e^f$,\ although 
perhaps not always\ (for the general Kasner case, there is one 
equation relating several scale factors\ $e^f,e^{h_m}$). For such a 
scale factor $e^f$ that is asymptotically $e^f\ra 1$,\ \eg\ \ 
$e^f=\tanh^a(x^+)$,\ one can in principle find a solution for\ $e^h$. 
Indeed an approximate solution of this kind (in the asymptotic region) 
is given\footnote{The exponent in the near-singularity form of $e^h$ 
may not be integral of course.} by eq.(\ref{asympsoln}). Choosing 
$e^f$ that is asymptotically $e^f\ra 1$, the spectrum of masses of 
string states are asymptotically as in flat space, while the near 
singularity spectrum is as discussed above.

Some oscillator states becoming increasingly massive is reminiscent of
\cite{crapsetal2} who argue for finite energy of free string propagation 
across plane-wave singularities. We note however that we have essentially 
analysed the free string spectrum in the vicinity of the singularity
in these backgrounds. Although formally it is possible to continue the
string mode expansion across the singularity, it would seem that the
physically relevant question would be to try and understand the role
of string interactions in the vicinity of the singularity, to obtain a
better understanding of string propagation across the singularity.

If string interactions generate a nontrivial (semiclassical) dilaton
profile say $\Phi(x)$ (that is regular), then presumably this is a way 
the background is desingularized, \eg\ if the backreacted background 
satisfies an equation of the form\ $R_{MN}\sim\del_M\Phi\del_N\Phi$. 
The solutions of these equations coincide with the singular 
background for $\Phi=0$ and are regular when a nonzero $\Phi$ is 
generated (although possibly string scale curvature).

Now we make a few comments on drawing insights into the AdS/CFT
cosmological investigations \cite{dmnt,adnnt} from our analysis here.
We have essentially used the scale factors $h_m(x^+)$ in our solutions
here to simulate the role of the dilaton there in that the internal
$h_m(x)$ scale factors shrinking effectively drive the singularity in
the $x^i$-directions, just as the time-varying dilaton drives the
singularity in the AdS/CFT cosmological context. It would then seem
that interaction effects between the various string modes could become
non-negligible near the null singularity in the bulk, although the
original classical bulk background might possess $\al'$-exactness
properties. This would be dual to possible nontrivial corrections to
the gauge theory effective potential stemming from loop effects, the
time-dependent gauge coupling being $g_{YM}^2=g_s=e^\Phi$. It would be 
interesting to explore this.

Finally it is interesting to ask if there are universal features in
the behaviour of string oscillator modes near generic time-dependent
singularities. For example, internal 6-dim spaces with intrinsic time
dependence, \eg\ due to closed string tachyon instabilities, will give
rise to 4D cosmological dynamics. Consider the case of unstable
noncompact conifold-like singularities \cite{knconiflips} embedded in
some compact space (say a nonsupersymmetric orbifold of a Calabi-Yau).
Phase diagrams obtained in the noncompact limit from appropriate
gauged linear sigma models show evolution from one of the two
classical phases corresponding to small resolutions to the other more
stable one through a flip transition \cite{orbflips,knconiflips},
involving the blowdown of a 2-cycle and a blowup of the topologically
distinct 2-cycle. From the point of view of the 4-dim effective field
theory, the sizes of these cycles are time-dependent scalars whose
spontaneous time evolution governs the 4-dim cosmology. In particular,
one might imagine that as we approach a flip singularity in the
internal space, a time-dependent 4-dim singularity develops. While a
direct stringy analysis of such a transition and resulting 4-dim
cosmology seems a priori difficult, it would be interesting to ask if
simple models using internal scale factors of the form studied here
can be used to mimic the internal time-dependence of
collapsing/growing cycles and study the resulting string dynamics,
possibly along the lines of \cite{hellermanswanson}.

\vspace{5mm}
%\newpage
\noindent {\small {\bf Acknowledgments:} It is a pleasure to thank 
S.~Das, R.~Gopakumar, S.~Govindarajan, P.~Mukhopadhyay, S.~Minwalla, 
and N.~Suryanarayana for useful discussions and conversations.
KN thanks S.~Das and S.~Trivedi for an enjoyable collaboration, some 
of the discussions from \cite{dmnt} in particular leading to this 
investigation. KM thanks the hospitality of CMI and IMSc over 
Nov-Dec '07 and Feb-May '08 respectively. We thank the organizers of 
the Monsoon Workshop, TIFR, Jun '08, and the ISM08 Workshop, 
Pondicherry, Dec '08, for hospitality and a stimulating environment. 
The work of KN is partially supported by a Ramanujan Fellowship, 
DST, Govt of India.}

%\vspace{5mm}

\appendix
\section{Some properties of the spacetime backgrounds}

\subsection{Lightcone supersymmetry of the backgrounds}

Here we analyse the supersymmetry of the Kasner-like backgrounds described 
here, although we have not really used this in our analysis in the paper. 
Choose the obvious diagonal orthonormal frame\ 
$e^+=e^{f/2}dx^+ ,\ e^-=e^{f/2}dx^- ,\ e^i=e^{f/2}dx^i ,\ e^m=e^{h_m/2}dx^m$ .\ 
The spin connection 1-forms are defined by\ 
$de^a + \omega{^a}_b\wedge e^b = 0$, where raising/lowering is performed by 
the flat space frame-metric. This gives the spin connection 1-forms to be
\bea
\omega_{-+} = -{1\over 2} f'dx^+\ , \qquad 
\omega_{+i} = -{1\over 2} f'dx^+\ , \qquad 
\omega_{+m} = -{1\over 2} h_m'dx^m\ ,
\eea
(It can be checked that these give the coordinate basis curvature 
components given previously.)
Taking the supersymmetry parameter $\epsilon$ to be a function only of the 
lightcone time, $\epsilon(x^+)$, the supersymmetry variation of the dilatino 
is trivially zero in this purely gravitational background with unexcited 
dilaton and RR/NSNS fluxes. The supersymmetry variation of the gravitino 
$\delta\psi_M$ (using eqs.(2.1,2.2) of \cite{granapolch} (see also 
\cite{schwarz})) reduces to
\be
D_M\epsilon = 
\left(\del_M + {1\over 8} \omega^{ab}_M [\Gamma_a,\Gamma_b]\right)\epsilon 
= 0\ ,
\ee
where $\Gamma_a$ are flat space $\Gamma$-matrices satisfying 
$\{\Gamma_a,\Gamma_b\}=2\eta_{ab}$, the curved space ones being
$\gamma_\mu=e_\mu^a\Gamma_a$. Taking $\epsilon$ to be $x^i,x^m$-independent 
is consistent with $D_M\epsilon=0, M\neq +$. Alongwith $D_+\epsilon=0$, 
this gives
\be
\Gamma^+\epsilon = 0\ , \qquad \left(\del_+-{1\over 4}f'\right)\epsilon = 0
\ee
which can be solved giving\ $\epsilon = e^{f/4} \eta$, where $\eta$ is a 
constant spinor satisfying $\Gamma^+\eta=0$ (it can be taken to be 
$\eta\sim\Gamma^+\chi$ where $\chi$ is some arbitrary constant spinor).
Closure of the algebra gives the equations of motion $R_{MN}=0$.
Thus these spacetime backgrounds preserve 16 real (lightcone) supercharges.

\subsection{Higher derivative curvature corrections}

As is often the case with lightlike backgrounds, these spacetimes do not 
appear to admit $\al'$ corrections due to higher order curvature terms. 
This is expected since these are, after a coordinate transformation, 
anisotropic plane-wave-like backgrounds (\ref{planewave}) which are
known to have such $\al'$-exactness properties \cite{horowitzsteif}. 
We outline below some rudimentary analysis of the vanishing of higher
derivative terms in the cosmological coordinates (\ref{absolns}),
(\ref{absolns2}), mainly for completeness.

At the level of the action, this is straightforward to see: with
$R_{++}$ alone being nonzero, there are no nonzero contractions since
there are no tensors with two or more upper $+$-components.
At the level of the equations of motion, one could ask if there are 
corrections to $R_{++}=0$ from higher order curvature terms. In this 
regard, various straightforward checks do in fact suggest the absence 
of corrections although we do not prove this in a theorematic way.

To elaborate a little, it is straightforward to see that no corrections 
of the form $f(R) R_{++}$ can arise where $f(R)$ is a complete 
contraction since the latter vanishes.
Let us therefore consider possible higher order terms of the form\
$A_{++}=R_{+M+N} T^{MN}=R_{+M+N} g^{MP} g^{NQ} T_{PQ}$, where $T_{PQ}$ is some 
tensor built out of $R_{MN}, R_{MNPQ}$ etc. Analysing the possible values 
for the indices forced by the contractions, we see that if $T_{PQ}$ has 
only $T_{++}$ nonzero, then $A_{++}$ vanishes since the background 
has\ $R_{+-+-}=0$. Thus \eg\ a possible correction at 
${\cal O}(R^2)$ of the form $R_{+M+N} R^{MN}$ vanishes. It is 
straightforward to further show that any correction $A_{++}$ with \eg\
$T_{PQ}\equiv R^{(k)}=R_{PP_1} {R^{P_1}}_{P_2} {R^{P_2}}_{P_3}\ldots {R^{P_k}}_Q$ 
vanishes:\ this can be seen by expanding\ $T_{PQ}\equiv R^{(k)}$\ to 
obtain the form\ $g^{P_1Q_1}g^{P_2Q_2}\ldots R_{PP_1}R_{Q_1P_2}R_{Q_2P_3}\ldots$, 
and noting that $g^{++}=0$ and $R_{++}$ alone is nonzero. Thus all higher 
order corrections with $T_{PQ}$ built from the Ricci tensor vanish.\\
Similarly, it is possible to show that a correction of the form \eg\ 
$R_{+M+N} R^{MPLQ} {R^N}_{PLQ}$ vanishes. It would seem that this would be 
possible to generalize to all orders as well.

The backgrounds in question have nonvanishing Weyl components $C_{+i+i} ,
C_{+m+m}$, with index structure as for $R_{ijkl}$. Thus higher derivative 
corrections involving the Weyl tensor are similar in structure and also 
vanish.

\section{An alternative time parameter and quantization} 

We have been working with $x^+$ as the time parameter so far. We will
now outline the analysis of this system with a canonical time parameter 
with $g_{+-}=-1$, and indicate results similar to the ones we have
discussed so far. Consider a coordinate transformation to the affine
parameter $\lambda$ as the time parameter transforming the metric
(\ref{absolns}) to
\be
ds^2 = -2d\lambda dx^- + \lambda^{a'} dx^idx^i + \lambda^{b'} dx^mdx^m\ ,
\ee
where\ $a'={a\over a+1} ,\ b'={b\over a+1}$.
Null congruences now have a natural time parameter here with\ 
$\xi={d\over d\lambda}$. Thus the geodesic deviation equation gives the 
acceleration norms as
\be
|a^i|^2 \sim {1\over\lambda^{4-a'}}\ , \qquad 
|a^m|^2 \sim {1\over\lambda^{4-b'}}\ ,
\ee
giving a singularity for\ $a',b'<4$, which are the same as the 
conditions (\ref{abCondns}).

Performing a lightcone gauge string quantization with $\tau\equiv\lambda$, 
we here obtain $E=-{1\over g_{+-}}=1$, so that in this case we effectively 
have conformal gauge also, as discussed earlier (see Sec.~3). Now the 
worldsheet action becomes
\be
S = {1\over 4\pi\al'} \int d^2\sigma\ \left( g_{II} (\del_\tau X^I)^2 
- g_{II} (\del_\sigma X^I)^2 \right)\ .
\ee
The equations of motion for the time-dependent modes $f^I_n$ now are\ 
$\del_\tau (\tau^{A_I}\del_\tau f^I_n) + n^2\tau^{A_I} f^I_n = 0$,\ 
where $A_I\equiv a',b'$. These give the mode functions
\be
f^I_n(\tau) = c^I_{n1} \sqrt{n} \tau^{{1-A_I\over 2}} 
J_{{A_I-1\over 2}}(n\tau) + c^I_{n2} \sqrt{n} \tau^{{1-A_I\over 2}} 
Y_{{A_I-1\over 2}}(n\tau)\ .
\ee
For low lying oscillation modes with finite $n\lesssim {1\over\lambda} $, 
these have the asymptotics\ $f^I_n\ \ra\ \ c^I_{n0} + c^I_{n\tau} \tau^{1-A_I}$ .
These string mode amplitudes thus do not diverge for $X^i_n$ since\ 
$a'={a\over a+1}<1$\ always, while the $X^m_n$ mode amplitudes normalized 
with the metric behave as\ $\tau^{b'}(X^m_n)^2$ which is finite if\ 
$b'<2$, \ie\ the same conditions ($b<2a+2$) as before.

The conjugate momenta are\ $\Pi^I={1\over 2\pi\al'} g_{II} (\del_\tau X^I)$. 
The Hamiltonian for this system (rewriting the $\Pi^I$ in terms of 
$\del_\tau X^I$), 
\be
H = {1\over 4\pi\al'} \int d\sigma\ \left( g_{II} (\del_\tau X^I)^2 
+ g_{II} (\del_\sigma X^I)^2 \right)\ 
\ee
is the generator of $\lambda$-translations, rather than $x^+$-translations. 
For the zero modes, the center-of-mass momenta are\ 
$p_{I0}(\tau)=\int d\sigma \Pi^I={1\over\al'} g_{II} {\dot X^I_0}(\tau)$.
Then the zero mode terms in the expression for the masses cancel between\ 
$2g^{+-}(-H_0)(p_-)-g^{II} (p_{I0})^2$. 

The oscillator contributions for the low lying modes can be 
calculated near the singularity as before, using their limiting 
expressions\ 
$f^I_n\ra c^I_{n0}\ ,\ {\dot f}^I_n\ra c^I_{n\tau} \tau^{-A_I}$.\\
Thus these low lying oscillator terms in the Hamiltonian above can again 
be simplified using (\ref{intermediateterms}) and rewritten in terms 
of one set of operators with coefficient\ 
$g_{II}({\dot f}^I_n)^2\ra \tau^{A_I}\tau^{-2A_I}=\tau^{-A_I}$,
and another set of operators with coefficient\ $g_{II}=\tau^{A_I}$.
This is identical in form to the expression for the masses (\ref{masses}) 
earlier, after resubstituting\ 
$A_I\equiv a',b'={a\over a+1} ,\ {b\over a+1}$ .

Now we consider the highly stringy modes: for any cutoff\ $\tau_\epsilon$, 
there are modes with\ 
$n\gg n_\epsilon={1\over\tau_\epsilon}={1\over\lambda}$\ whose asymptotics 
is essentially like a plane-wave, with\ 
$f^I_n\ra \tau^{-{A_I\over 2}} e^{in\tau}$.\ Using these, the Hamiltonian 
above simplifies using (\ref{intermediateterms}) to
\bea
H_\lambda &\sim&
{1\over \al'} \sum_n {1\over n^2} \Biggl( ( a^I_{-n}a^I_n + 
{\tilde a}^I_{-n}{\tilde a}^I_n + n ) ( g_{II} |{\dot f}^I_n|^2 
+ n^2 g_{II} |f^i_n|^2 ) \nonumber\\
&& \qquad\qquad -\ a^I_n {\tilde a}^I_n ( g_{II} ({\dot f}^I_n)^2 + n^2 
g_{II} (f^i_n)^2 ) - a^I_{-n} {\tilde a}^I_{-n} ( ({\dot f}^{I*}_n)^2 + 
n^2 g_{II} (f^{I*}_n)^2 ) \Biggr)\ .
\eea
Using the $f^I_n$ asymptotics, the terms in the second line are 
vanishingly small while the $\tau$-dependent terms in the first line 
are\ $\tau^{A_I} (2n^2) \tau^{-A_I}$.\ This Hamiltonian has the same form 
as $g^{+-} H_{x^+}$, using the expression (\ref{Hlargen}) for $H_{x^+}$, 
for the highly stringy modes near the singularity.

We now describe some aspects of string quantization in the Brinkman 
coordinates (\ref{planewave}), after redefining to the affine parameter 
$\lambda$. The metric is\ 
$ds^2=-2d\lambda dy^-+\sum_I\chi_I(y^I)^2{d\lambda^2\over\lambda^2}+(dy^I)^2$,
with\ $\chi_I=\frac{a_I}{2(a+1)}(\frac{a_I}{2(a+1)}-1)$.\ 
The string action is\ 
$S={1\over 4\pi\al'} \int d^2\sigma\ ( (\del_\tau y^I)^2-(\del_\sigma y^I)^2
+\sum_I{\chi_I\over\tau^2}(y^I)^2 )$.\ The equations of motion give the 
mode functions
\be
f^I_n(\tau)=\sqrt{n\tau} (c^I_{n1} J_{{\sqrt{1+4\chi_I}\over 2}}(n\tau) + 
c^I_{n2}Y_{{\sqrt{1+4\chi_I}\over 2}}(n\tau))\ , 
\ee
resulting in a mode expansion similar to (\ref{modeexpXIn}), with\ 
$k^I_n={i\over n}\sqrt{{\pi\al'\over 4}}$ .
The highly stringy modes are defined by the limit of small $\tau$, large $n$, 
and $n\tau\gg 1$. Then\ $f^I_n\sim\ e^{-in\tau}$ for $c^I_{n1}=1, c^I_{n2}=-i$, 
and the Hamiltonian from the action above reduces to\
\be
H \sim\ \sum_{n\gg 1/\tau}\ (1-{\chi_I\over 2n^2\tau^2}) 
(a^I_{-n}a^I_n+{\tilde a}^I_{-n}{\tilde a}^I_n+n) - 
{\chi_I\pi\over 4n^2\tau^2} (a^I_n{\tilde a}^I_n (f^I_n)^2 + 
a^I_{-n}{\tilde a}^I_{-n} (f^{I*}_n)^2)\ .
\ee
Thus the Hamiltonian for the highly stringy modes exhibits similar 
behaviour here as earlier\footnote{Similar expressions arise from the 
corresponding limit in \cite{prt} (sec.~6).}. Similarly, defining the 
$b^I$-oscillators as before, the Hamiltonian for the low-lying 
oscillator modes is\ $H\sim\ \sum_{n\lesssim 1/\tau} {\pi\over 4n^2}\ 
(b^{I\dag}_{n\tau}b^{I}_{n\tau}+(n^2-{\chi_I\over\tau^2})b^{I\dag}_{n0}b^{I}_{n0})$.

\end{document}